©SHUTTERSTOCK.COM/LUCKYSTEP

# 5G Channel Models for Railway Use Cases at mmWave Band and the Path Towards Terahertz

**Ke Guan\*,**
*Senior Member, IEEE.*
E-mail: kguan@bjtu.edu.cn

**Juan Moreno,**
*Senior Member, IEEE.*
E-mail: juanmorenogl@diac.upm.es

**Bo Ai,**
*Senior Member, IEEE.*
E-mail: boai@bjtu.edu.cn

**Cesar Briso-Rodriguez**
*Member, IEEE.*
E-mail: cesar.briso@upm.es

**Bile Peng,**
*Member, IEEE.*
E-mail: bile.peng@chalmers.se

**Danping He,**
*Member, IEEE.*
E-mail: hedanping@bjtu.edu.cn

**Andrej Hrovat,**
*Member, IEEE.*
E-mail: andrej.hrovat@ijs.si

**Zhangdui Zhong,**
*Senior Member, IEEE.*
E-mail: zhdzhong@bjtu.edu.cn

**Thomas Kürner**
*Senior Member, IEEE.*
E-mail: kuerner@ifn.ing.tu-bs.de

\*Corresponding author

*Abstract*—High-speed trains are one of the most relevant scenarios for the fifth-generation (5G) mobile communications and the "smart rail mobility" vision, where a high-data-rate wireless connectivity with up to several GHz bandwidths will be required. This is a strong motivation for the exploration of millimeter wave (mmWave) band. In this article, we identify the main challenges and make progress towards realistic 5G mmWave channel models for railway use cases. In order to cope with the challenge of including the railway features in the channel models, we define reference scenarios to help the parameterization of channel models for railway use at mmWave band. Simulations and the subsequent measurements used to validate the model reflect the detailed influence of railway objects and the accuracy of the simulations. Finally, we point out the future directions towards the full version of the smart rail mobility which will be powered by terahertz (THz) communications.

## I. Introduction

The upcoming fifth-generation (5G) mobile communication system is expected to support high mobility up to 500 km/h, which is envisioned in particular for high-speed trains (HST). Millimeter wave (mmWave) spectrum is considered as a key enabler for offering the "best experience" to users moving at high speeds. Apart from the requirement of 5G, rail traffic itself is expected to evolve into a new era of "smart rail mobility" where infrastructure, trains, passengers and goods will be fully interconnected, which means also more complicated use cases. To realize this vision, which is one part of the "Shift2Rail Initiative" [1], the European Union is calling proposals on topics like intelligent rail infrastructure, intelligent mobility management, smart rail services, etc. Indeed, these requirements impose requirements for a seamless high-data-rate wireless connectivity in railways.

Correspondingly, railway communications are evolving from only supporting critical signaling applications to providing several bandwidth-intensive applications, like onboard and wayside high definition (HD) video surveillance, high-data rate connectivity for passengers, passenger information, remote driving, among others. These applications need to be deployed in at least five scenarios: train-to-infrastructure (T2I), inter-car (the train backbone), intra-car, inside station, and infrastructure-to-infrastructure [2]. For both the station and infrastructure-to-infrastructure scenarios, the bandwidth requirements go from several hundred MHz to several GHz. For sure, the biggest challenge is posed by the T2I scenario because this link needs to achieve very high data rates, low latencies, as well as close to 100% availability while traveling at speeds around 500 km/h (as of 2018, high-speed trains do not exceed 350 km/h in commercial travels, but the speed record is at 574 km/h since 2007). As the main interface between the network on-train and the fixed network, in the T2I scenario an aggregated stream for the backhaul for both inter/intra-car scenarios is transmitted. Therefore, a bandwidth of several GHz (or even higher) is needed to accommodate up to 100 Gbps data rates. Such high-data rate and huge bandwidth requirements are a strong motivation to explore the mmWave band [3], [4]. This exploration means that we need to have a great understanding of the behavior of the channel at mmWave band in railway scenarios.

Regarding mmWave bands for 5G, the standardization race has started to gain considerable traction within both the mobile industry and the European Union, through the 5G Public-Private Partnership (5GPPP) initiative [5]. In particular, many countries have started to test 5G-related technologies for railway use in mmWave band. For example, in Germany, some preliminary experiments were carried out at 35 and 58 GHz bands; in Japan, 36–40.5 GHz, 43.5–52.6 GHz, 55.78–76 GHz, 92–94 GHz, 94.1–100 GHz, and 102–109.5 GHz are the main concerned bands for T2I; whereas the 60 GHz band is of great interest for French railways, 67 GHz in the UK and 26 GHz in Korea, to provide only some examples.

Therefore, it is critical to have fundamental knowledge of the propagation channel characteristics in those bands. However, from the wireless channels point of view, there are numerous open challenges [6] on researching and developing 5G mmWave systems for railway communications. To summarize, the sparseness of systematic research leads to the lack of standard channel models and reference scenarios in railways for link-level or system-level evaluation. Therefore, it is of primary importance to have standardized channel models in order to do properly all the application layer end-to-end tests for all the relevant use cases that will be enabled by these high-capacity communication systems in railways.

The rest of this article is organized as follows. In Section II, we discuss the challenges and solutions for involving railway features in channel models for mmWave band. Moreover, two versions of comprehensive railway scenario modules are reconstructed with detailed geometry and material information to highlight how the modeling could be. To do so, two stochastic channel models

> It is of primary importance to have standardized channel models in order to do properly all the application layer end-to-end tests for all the relevant use cases that will be enabled by these high-capacity communication systems in railways.

in two railway scenarios are established and validated at 90 GHz and 30 GHz, respectively. Future directions and a roadmap of realization of the full version of the smart rail mobility are pointed out in Section III. Conclusions and future work are drawn in Section IV.

## II. Involving Railway Features in 5G mmWave Channel Models

In order to evaluate 5G mmWave communication paradigms in railway environments, it is critical to define reference scenarios with typical railway propagation features [7]. However, existing mmWave channel models are mainly for standard environments, e.g., Winner II [8], where railways are treated as a particular use case of a "moving hotspot". Therefore, including the railway features in channel models is still an open challenge. Without the consideration of the influence of the main objects and geometries in propagation environments, the prototypes that can work in the lab may become unpredictable and unreliable in the real implementations. In this study, we present two versions of the reference scenario modules: the complete version with all the objects and the concise version with only the most significant objects. Based on the ray-tracing simulations in these scenario modules, 5G channel characterization and modeling that reflect the railway features can be realized.

### A. Complete Version of Reference Scenario Modules of mmWave and THz Railway Channels

In order to provide a baseline for evaluating 5G mmWave communications for railways, the commonness of typical railway scenarios should be abstracted. Thus, rather than modeling the site-specific railway environments one by one, we reconstruct the three-dimensional (3D) model of a set of comprehensive railway scenarios including six modules for 5G mmWave railway channels. As shown in Fig. 1, they are defined as "Module 1– Tunnel entrance on steep wall connecting cutting with crossing bridges," "Module 2– Viaduct with open train station," "Module 3 – Urban with semi-closed train station," "Module 4–Rural with cut and cover tunnel," "Module 5–Rural connecting double-track tunnel," and "Module 6–Single-track viaduct." All the main objects are modeled according to the typical geometries and materials in reality (see Table I for more details). Apart from common objects, such as trains, tracks, pylons, traffic signs, billboards, etc., each module includes its special objects, like steep walls, cutting walls, crossing bridges, train stations, indicators, barriers, vegetation, cut and cover tunnels, dual-track tunnel, and single-track viaduct. All these scenario modules can be independently used for concept verification in site-specific mmWave railway scenarios, or they can be combined in various ways for doing statistic evaluation of the system. The 3D models of the six modules are publicly available and freely downloadable for link-level and system-level simulations (http://raytracer.cloud).

Table I. Parameters of complete version of railway scenarios.

| Relative Permittivity, Loss Tangent | Objects |
|---|---|
| Metal: 1.00, $10^7$ | Billboard, barrier, traffic signs, crossing bridge, pylon, awning |
| Concrete: 1.06, 0.65 | Pylon of billboard, track, tunnel, buildings in rural, steep wall, CCT, cutting wall, ground, station base |
| Aluminium alloy: 1.29, $10^7$ | High speed train |
| LED: 3.74, 3.14 | Indicator, billboard |
| Tempered glass: 10.00, 0.43 | Buildings in urban |
| Vegetation: 29.12, 0.278 | Vegetation |
| Smooth marble: 1.96, 0.30 | Buildings in urban |
| Ceramic tile: 1.85, 0.07 | Buildings in urban, platform and pylons in station |

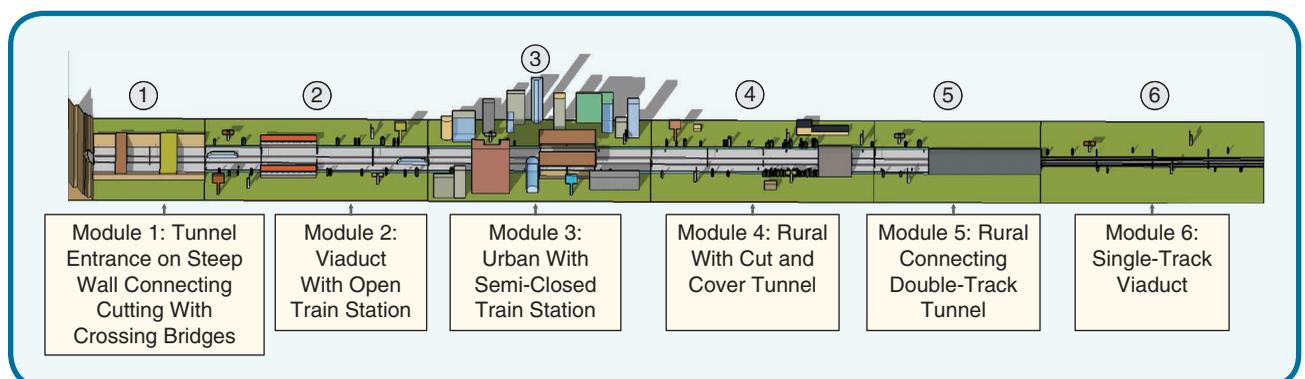

**FIG 1** Complete version of railway scenarios including six modules for 5G mmWave railway channels, originating from [7].

> The 3D models of the six modules are publicly available and freely downloadable for link-level and system-level simulations (http://raytracer.cloud).

*B. Channel Characterization in Complete Version of Modules*

In this study, we use a self-developed broadband and dynamic channel ray-tracing simulator for channel simulation–CloudRT [9]. It combines a railway ray-tracing simulator [10] and a broadband THz ray-tracing simulator [2], [11]. The former is validated by extensive measurements in [12]–[14], while the latter is calibrated and validated by a large number of measurements in [11]. Recently, CloudRT has been updated to a high-performance computing platform with the website (http://raytracer.cloud).

In order to characterize the 5G mmWave railway channels, ray-tracing simulations with two antenna height setups in the 60 GHz band with 8 GHz bandwidth are made in all six modules. The simulations are based on frequency domain ray-tracing principle for ultra-wideband signal, and the electromagnetic properties of materials are either from measurements or literature. The setup parameters are shown in Table II. The propagation mechanisms considered in the simulations are line of sight (LOS), scattering (multiple scattering theory for vegetation and single lobe directive model for others), diffraction (Uniform Theory of Diffraction), and reflection.

We define two antenna setups: both the transmitter (Tx) and the receiver (Rx) in the setup 1 (6 m and 4.5 m) are higher than those in the setup 2 (1 m and 0.92 m). It is also noteworthy that in this study, we do not need to add directive antennas into the simulation because the radiation pattern could be included in the post-processing to analyze the influence of various antenna patterns or beamforming strategies.

The power delay profile (PDP), root-mean-square (RMS) delay spread, path loss exponent, shadow factor, Ricean K-factor, RMS delay spread, and coherence bandwidth were analyzed in our previous publication [15]. These channel characteristics show that the objects that are not so relevant to sub-6 GHz channels indeed influence mmWave channels.

In this study, we move on to the angular domain and analyze the four angular spreads – azimuth angular spread of arrival (ASA) and departure (ASD), elevation angular spread of arrival (ESA) and departure (ESD). They jointly define the distribution of the departure- and arrival angles of each multipath component in 3D space seen by the Tx and Rx, respectively. Therefore, these spreads can be translated into specific angles for each multipath cluster, serving as key parameters in various channel models, such as the whole measurement-based pseudo-geometric model (MBPGM) families [16]–WINNER-style (like 3GPP spatial channel model (SCM), extended SCM (SCME) [17], WINNER (WIM1), WINNER II (WIM2) [18], [19], WINNER+ (WIM+) [20], 3D multiple input and multiple output (MIMO) channel models [21], [22], and QuaDRiGa [23]) and COST-style (e.g., COST2100 [24])–as well as the geometry-based stochastic models (GBSM) families [25], [26].

In this study, the four angular spreads are calculated in the usual way [27]. Moreover, when jointly analyzing the angular spreads and the PDP in each scenario, we can see clearly how the railway objects influence the channel in both time and angular domains. Generally speaking, (also can be seen in the particular scenario that is depicted in Fig. 2, ASD and ASA are larger than ESD and ESA, which implies that most of the multipath components come from the horizontal direction. This truthfully reflects the fact that more objects (scatterers) are located on the two sides of railway rather than over or under the train. Buildings, ground, billboards, pylons, and vegetation generate effective multipath components either independently (single reflection,

| Table II. Simulation setup. | |
|---|---|
| Scenarios | The Six Modules |
| Antenna type and gain | Omni-directive, 0 dBi |
| Location of Tx and Rx | Tx: pylon, Rx: windshield of train |
| Heights of Tx and Rx | Setup 1: 6 m and 4.5 m |
| | Setup 2: 1 m and 0.92 m |
| Mobility model and speed of the Rx train | Rx moving away from Tx along the track at a constant speed of 500 km/h |
| System bandwidth | 8 GHz: 60.32 GHz - 68.32 GHz |
| Frequency points | 801 |
| Transmitting power | 0 dBm |
| Sample interval | 2 mm |

> Apart from the channel characterization in the complete version of the modules, another solution is to establish channel models directly based on extensive ray-tracing simulations.

scattering, diffraction) or jointly (multiple reflections). These multipath components are strongly time-variant, and in particular, the angular spreads increase tempestuously when these multipath components are presented. As an example shown in Fig. 2, because of the low antenna setup (antenna setup 2 Tx and Rx are 1 m and 0.92 m high), the multipath components in the vertical direction are sparse, and therefore, the ESA and ESD are only around 10° (with occasional jumps caused by sudden appearance of strong multipath components). Also because of the very low antenna heights, more objects in the horizontal direction – such as train station base, barrier, and moving train – which are insignificant in the high-antenna channel do influence the low-antenna channel. On one hand, their appearance leads to high ASD and ASA (which gradually decrease when Rx approaches Tx); on the other hand, their sudden absence causes temporal slumps in ASD as well (see the black circles in Fig. 2). To summarize, through ray-tracing simulations in the proposed six railway scenario modules, all the fundamental channel characteristics at mmWave band including corresponding parameters of the path loss exponent, shadow factor, Ricean K-factor, RMS delay spread, coherence bandwidth, and the four angular spreads (ASA/ASD/ESA/ESD), can be obtained and analyzed. These parameters, reflecting the 5G mmWave railway channel features, can be input into various channel models. Similar methods that also make use of ray-tracing results to parameterize the urban channel in the 5G mmWave standard channel models can be found in [28].

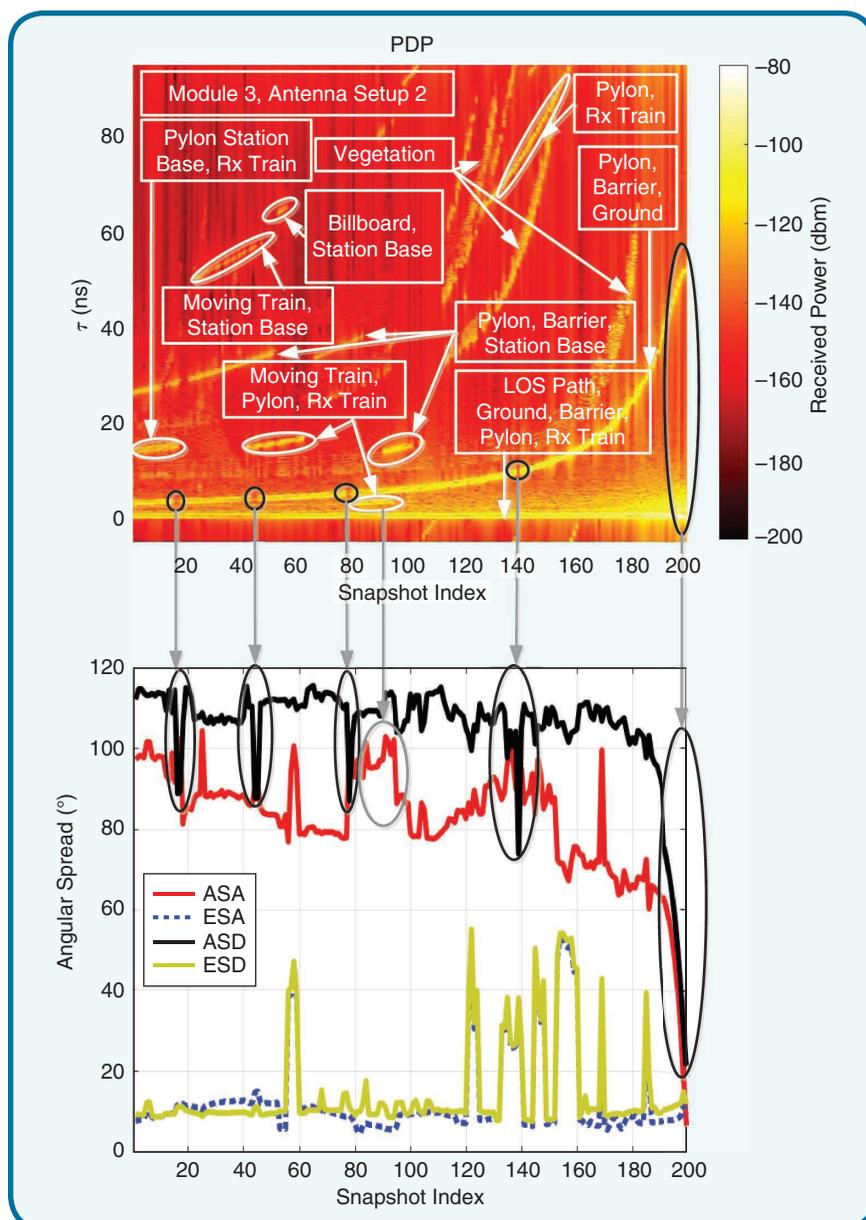

**FIG 2** PDP and angular spreads of Module 3 with antenna setup 2. In the bottom figure, the ASA, ASD, ESA, and ESD.

### C. Channel Models in the Concise Version of Scenarios

Apart from the channel characterization in the complete version of the modules, another solution is to establish channel models directly based on extensive ray-tracing simulations. In this regard, a good tradeoff between the accuracy and the conciseness of the environment reconstruction is required. Thus, we present the concise version of the railway environment model only with the objects which have a significant contribution to mmWave channels. According to the previous significance analysis conclusions in our previous study [12], two concise railway scenarios are reconstructed for the mmWave band. The first scenario is for the outdoor railway environment: when Tx and Rx are lower than the barriers, the traction pylons, tracks and the objects outside the barriers are neglected; while when either Tx or Rx is higher than the barriers, the objects outside the barriers are reconstructed. The second scenario is for tunnels, which is modelled with the train, ground and tunnel walls, omitting tracks and tunnel furniture whose area is smaller than 7 m². The concise versions of the outdoor railway and tunnel scenarios are shown in Fig. 3. Generally speaking, compared with the complete version, the number of surfaces in the concise version decreases from 54.46% up to 78.71%. Correspondingly, the ray-tracing simulation time decreases from 69.95% up to 89.90% of that of the completed version.

> A good tradeoff between the accuracy and the conciseness of the environment reconstruction is required.

*1) Stochastic Channel Model Based on Ray-Tracing in the Concise Outdoor Railway Scenario at 90 GHz*

In the concise outdoor railway scenario, the same self-developed broadband and dynamic channel ray-tracing simulator that is used in Section II B is employed to do extensive simulations at center frequency 93.2 GHz with 2 GHz bandwidth. Following the 3GPP mmWave HST network architecture, the time-varying channel is generated according to the trackside Tx setting on the remote radio head (RRH) and a mobile relay station on the train in movement. The Tx is equipped with a vertically polarized omnidirectional antenna that is 3.2 m above the ground. The Rx fixed on the train's front window, 3 m above the ground, is omnidirectional and vertically polarized. With a sampling interval of 3.6 ms, the Rx moves toward the 580 m

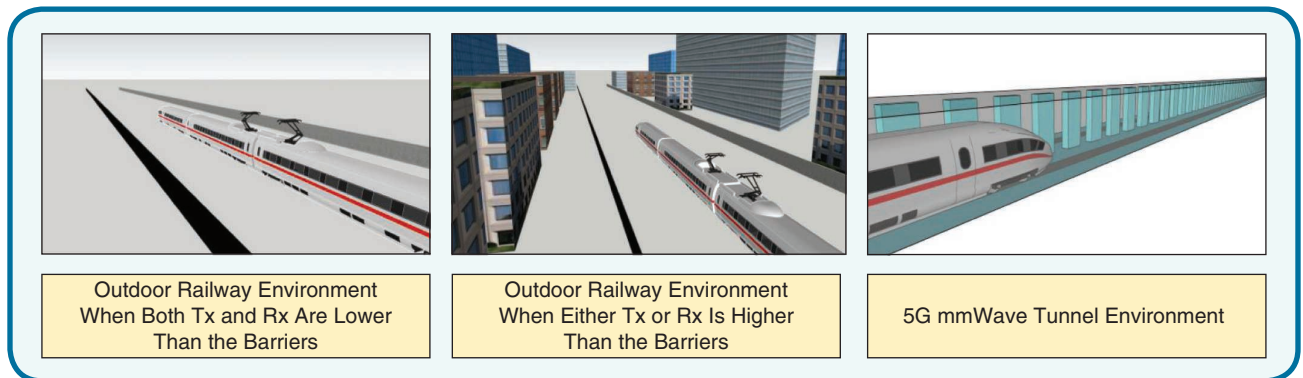

Outdoor Railway Environment When Both Tx and Rx Are Lower Than the Barriers

Outdoor Railway Environment When Either Tx or Rx Is Higher Than the Barriers

5G mmWave Tunnel Environment

**FIG 3** Concise version of the outdoor railway and tunnel scenarios.

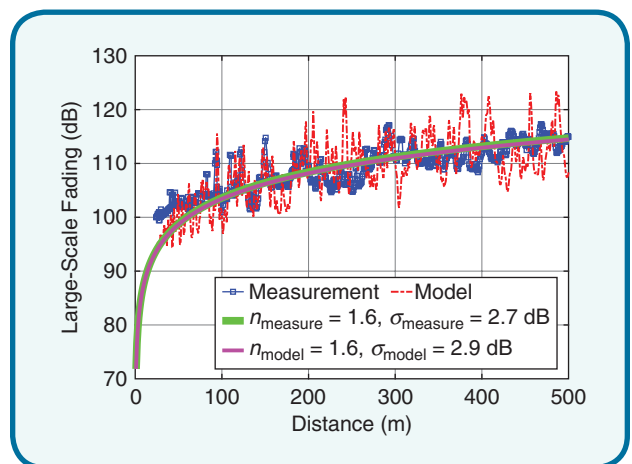

**FIG 4** Comparison of large-scale fading between measurement and model.

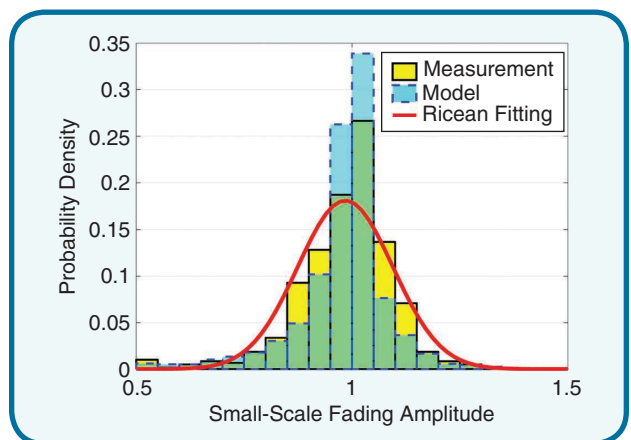

**FIG 5** Comparison of the PDF of small-scale fading between measurement and model.

> The proposed model based on ray-tracing in the concise outdoor railway scenarios can accurately reproduce the large-scale fading for link budget and the small-scale fading for evaluating and optimizing the transmission schemes enabling smart rail mobility and 5G railway use.

distant RRH at a speed of v = 500 km/h. Considering that the significant materials in this scenario – concrete, metal, tempered glass, resin – are not very frequency dispersive between 90 GHz and 100 GHz, we use the standard electromagnetic (EM) property of them at 100 GHz given by ITU-R [29]. Since the LOS always exists in this scenario, multiple orders of reflected rays and single-bounced scattered rays are traced as multipath components, while diffraction is not involved. Based on extensive ray-tracing simulations, a stochastic model is established. By implementing the formulas in [30], the complete channel transfer function (CTF) can be obtained. To validate this stochastic model, we employ measurements at 90 GHz band in the outdoor railway environment taken by the Railway Technical Research Institute (RTRI) and the National Institute of Information and Communication Technology (NICT) [31]. In order to separate the large-scale fading and the small-scale fading, a sliding/overlapped window with an interval of 10 wavelengths and a window size of 20 wavelengths for averaging are used for both the model results and the measurement data. As shown in Fig. 4, the measurement and the model have good agreements in terms of the large-scale fading. The path loss exponent fitted by model and measurement, respectively, is the same $-1.6$ ($n_{\text{measure}}$ and $n_{\text{model}}$); the shadow factor of model and measurement is 2.9 dB $\sigma_{\text{model}}$ and 2.7 dB $\sigma_{\text{measure}}$, respectively. The comparison of the probability density function (PDF) of the amplitudes of the small-scale fading is shown in Fig. 5, where the model result matches the measurement well, and it is found that the Ricean distribution can well characterize the amplitude distribution of the small-scale fading due to the presence of the LOS ray in the analyzed scenario.

The overall result of this validation can be summarized stating that there are good agreements on the measured and simulated large-scale fading and small-scale fading. To summarize, the proposed model based on ray-tracing in the concise outdoor railway scenarios can accurately reproduce the large-scale fading for link budget and the small-scale fading for evaluating and optimizing the transmission schemes enabling smart rail mobility and 5G railway use.

2) Stochastic Channel Model Based on Ray-Tracing in the Concise Tunnel Scenario at 30 GHz

In the concise tunnel scenario, using the same ray-tracing simulator, we conducted simulations at the center frequency 30 GHz with 500 MHz bandwidth. With the Rx speed of 500 km/h, the total number of snapshots are 500,000 in the target straight tunnel. All the details for this model can be seen in [14]. After evaluating the contribution of each propagation mechanism, a stochastic channel modeling for T2I link is established based on the LOS ray and up to 10th order of reflected rays; whereas the other rays (diffraction, scattering, and higher order reflections) are omitted due to their insignificant contribution. In order to get full polarization information, all the four polarization combinations are simulated (vertical-vertical, vertical-horizontal, horizontal-vertical and horizontal-horizontal). As we did in the outdoor model, the validation is carried out using measurements in a real scenario. In this case this scenario is a Seoul subway tunnel between 31.5–31.75 GHz with 250 MHz bandwidth (measurements done by ETRI [32]). As can be seen in Fig. 6, the comparison of signal-to-noise ratio (SNR) between the measurements and the stochastic model in the concise tunnel scenario shows a good agreement, not only in terms of propagation loss but also regarding fading (the mean error of the SNR is smaller than 1 dB and the standard deviation is smaller than 4 dB).

To summarize, the above channel models are two case studies. They demonstrate how to generate the realistic 5G channel models for railway use cases at mmWave band based on ray-tracing in the concise version of the railway environment models. Both the ray-tracing simulator and the railway environment models are open access at

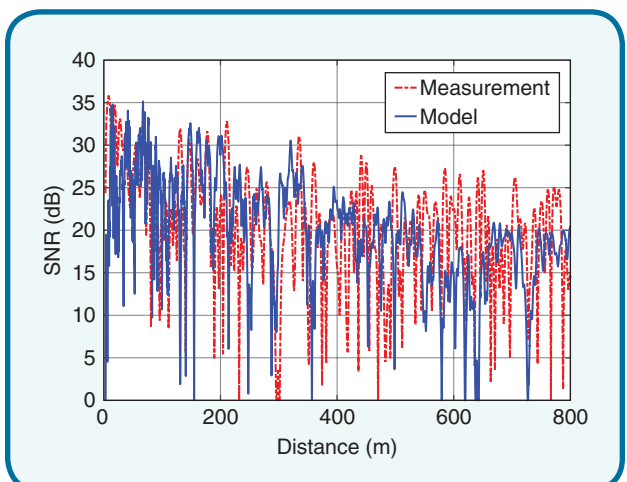

**FIG 6** Comparison of SNR between measurement in Seoul subway and stochastic model.

http://raytracer.cloud. The interested readers can easily utilize this resource for free to run simulations and obtain realistic channels in their target scenarios and frequency bands according to their preference.

## III. Future Directions on Smart Rail Mobility

In this section, we draw the roadmap and discuss the future directions by defining two versions of the smart rail mobility paradigm–the basic version and the full version–in terms of data rate, bandwidth, carrier frequency, technical challenges, and implementation timeline.

In the basic version, considering the available bandwidths at mmWave band from 1 GHz (at 28 GHz) to 9 GHz (at 60 GHz), data rates of 1–10 Gbps could be achieved even with moderate spectral efficiencies. However, considering adaptive beamforming at mmWave band, if using current beam training protocols involving an exhaustive search or prioritized sector search ordering, challenges will arise on how to keep the T2I link configuration time within the latency bounds of HD video or VoIP services. Moreover, novel handover scheme (e.g., efforts in [33]) and multi-beam concurrent transmission scheme (such as the heterogeneous multi-beam cloud radio access network architecture proposed in [34]) will be needed for improving handover performance and providing seamless mobility and coverage for mmWave network. To sum up, referring to the timeline of the standardization of 5G mmWave railway communications, e.g., IEEE 802.15 Interest Group HRRC (Interest Group High Rate Rail Communications), as well as the 5G standardization in 3GPP and ITU-R, mmWave communications are practical to enable the basic version of smart rail mobility in the next five years – 2020–2025.

For the full version, taking advantage of the ultra-high bandwidths beyond 20 GHz available in the terahertz (THz) frequency range, higher-data-rate transmission (i.e., beyond 100 Gbps) is possible to achieve. However, compared to the mmWave band, the beam width at THz frequencies is much narrower (even 1°) in order to compensate the higher path-loss. Thus, it is more challenging to realize adaptive beamforming in the T2I scenario. But, if we can accurately predict the train location, it is possible to avoid (or at least greatly mitigate) the need of beam training and then the aforementioned disadvantage of longer link configuration time at THz can be neglected. Note that the speed of a train can be time-variant, which will raise challenges to design competent localization algorithms. For inter-car and infrastructure-to-infrastructure scenarios, to overcome the performance loss from aerodynamics-induced beam misalignment, it is still necessary to develop efficient beam alignment techniques (like the new angle of arrival estimation algorithms in [35]), which are more challenging for THz due to the narrower beam widths. But from a long-term point of view, THz communications have broader prospects, because the very strong distance-frequency dependent characteristics of the THz band can motivate the development of novel communication schemes, such as, distance-aware multi-carrier modulation (DAMC), hybrid beamforming scheme with DAMC transmission, etc. To summarize, according to the progress on THz communication standardization, for instance, IEEE 802.15.3d-2017 and IEEE 802.15 IGthz, the full version of smart rail mobility providing a seamless high-data-rate connectivity beyond 100 Gbps is expected to be enabled by THz communications between 2025 and 2030.

> The full version of smart rail mobility providing a seamless high-data-rate connectivity beyond 100 Gbps is expected to be enabled by THz communications between 2025 and 2030.

## IV. Conclusion and Future Work

In this article, we identify the main challenges and present a series of work including railway scenario modeling, ray-tracing simulations, and channel modeling, towards realistic 5G channel models for railway use cases at mmWave band. We have seen that including railway features in channel models is a challenging task mostly because there are no channel models focused exclusively on railways. A possible solution for this is to use the standard railway scenarios for the ray tracer. Regarding this tool, we have shown the fundamental characteristics of the channel at the 60 GHz band that could serve as an input to various standard channel models in mmWave band. Furthermore, based on the ray-tracing simulations in both the outdoor railway and tunnel scenarios, two stochastic channel models were established and validated by measurements at 90 GHz and 30 GHz bands, respectively. The channel impulse responses generated by these models can be an input to link-level simulators to evaluate the physical layer technologies. However, currently there is as yet no widely accepted solution to verify the ray-tracing data in system-level modeling or to clarify how the propagation characteristics directly influence the system performance. Thus, future efforts should be made to link the propagation channel model with metrics such as path loss, shadow fading, delay spread, angular spreads, etc., to the system-level metrics, such as Precoding Matrix Indicators (PMI), Channel Quality Information (CQI), and so on.

Last but not least, future directions were pointed out as well. If the main technical challenges can be well addressed, the basic version of smart rail mobility enabled by 5G mmWave communications is very promising to come true by 2025, and a data rate between 1-10 Gbps could be expected. Then, in the subsequent 5-10 years, THz communications in the era of beyond 5G (B5G) could be the key enabler of the full version of smart rail mobility providing a seamless high-data-rate connectivity beyond 100 Gbps.

## Acknowledgment

This work is supported by the National Key Research and Development Program under Grant 2016YFE0200900, Beijing Natural Science Foundation under Grant L161009, NSFC under Grant (61771036 and 61725101), Fundamental Research Funds for the Central Universities 2018RC030, and Alexander von Humboldt Foundation.

## About the Authors

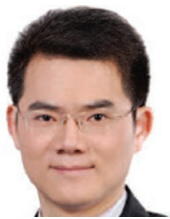
*Ke Guan*, received the Ph.D. degree from School of Electronic and Information Engineering of Beijing Jiaotong University in 2014. He is an Associate Professor in State Key Laboratory of Rail Traffic Control and Safety, Beijing Jiaotong University. His current research interests are in the field of measurement and modeling of wireless propagation channels for 5G, mmWave, THz, and intelligent transportation systems. E-mail: kguan@bjtu.edu.cn

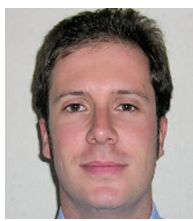
*Juan Moreno*, received the Ph.D. degree from the Universidad Politecnica de Madrid in 2015. He works as a rolling stock engineer in the Engineering and Research Department of Madrid Metro and as a part-time professor in the Universidad Politecnica de Madrid. His research interests are channel measurement & modelling, railway communications systems and software-defined radio. E-mail: juanmorenogl@diac.upm.es

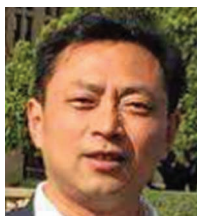
*Bo Ai*, received the Ph.D. degree from Xidian University in 2004 in China. He is now working in Beijing Jiaotong University as a professor and advisor of Ph.D. candidates. He is a deputy director of State Key Lab of Rail Traffic Control and Safety. His current interests are the research and applications OFDM techniques, HPA linearization techniques, radio propagation and channel modeling, GSM for railway systems, and LTE for railway systems. He is an IET Fellow and an IEEE Senior member. E-mail: boai@bjtu.edu.cn

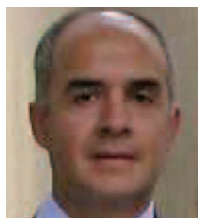
*Cesar Briso-Rodriguez*, received the Ph.D. degrees in Telecommunications Engineering from Universidad Politecnica de Madrid (UPM), Spain, in 1999. Since 1996, he has been Full Professor of the Telecommunication school of the Technical University of Madrid. His research activities have been focused on design and development of high frequency communication systems for railway. E-mail: cesar.briso@upm.es

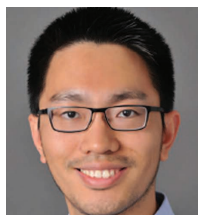
*Bile Peng*, received the Ph.D. degree from the Institut fuer Nachrichtentechnik, Technische Universitaet Braunschweig in 2018. He is currently a Project Assistant with the Chalmers University of Technology, Sweden. His research interests include wireless channel measurement, modeling and estimation, high precision vehicle localization and mapping, as well as artificial intelligence. E-mail: bile.peng@chalmers.se

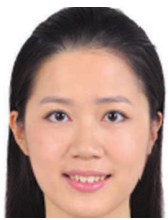
*Danping He*, received the Ph.D. degree from Universidad Politecnica de Madrid in 2014. She is currently conducting postdoctoral research in the State Key Laboratory of Rail Traffic Control and Safety, Beijing Jiaotong University. She is an Associate Professor in State Key Laboratory of Rail Traffic Control and Safety, Beijing Jiaotong University. Her current research interests include radio propagation and channel modeling, ray tracing simulator development and wireless communication algorithm design. E-mail: hedanping@bjtu.edu.cn

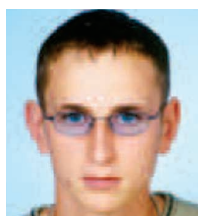
*Andrej Hrovat*, received the Ph.D. degree in Electrical Engineering from the Joef Stefan International Postgraduate School, Slovenia, in 2011. He is currently a researcher in the Department of Communication Systems of the Joef Stefan Institute and assistant at the Joef Stefan International Postgraduate School. His research interests include radio signal propagation, channel modeling, terrestrial and satellite fixed and mobile wireless communications, radio signal measurements and emergency communications. E-mail: andrej.hrovat@ijs.si

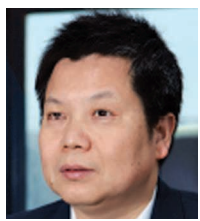
*Zhangdui Zhong*, received the Master's degree from Beijing Jiaotong University, Beijing, China, in 1988. He is a professor and advisor of Ph.D. candidates in Beijing Jiaotong University. He is now a Chief Scientist of State Key Laboratory of Rail Traffic Control and Safety in Beijing Jiaotong University. His research interests include railway communications, AI, big data, and cloud computing. E-mail: zhdzhong@bjtu.edu.cn

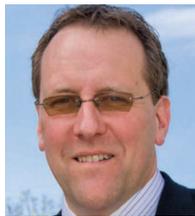

*Thomas Kürner*, received his Dr.-Ing. degree in 1993 from University of Karlsruhe (Germany). Since 2003 he is Full University Professor for Mobile Radio Systems at the Technische Universitaet Braunschweig. His working areas are indoor channel characterization and system simulations for high-speed short-range systems including future terahertz communication system, propagation, traffic and mobility models for automatic planning and self-organization of mobile radio networks, car-to-car communications as well as accuracy of satellite navigation systems. E-mail: kuerner@ifn.ing.tu-bs.de

## References

[1] "Shift2Rail: Driving innovation on railways," European Commission, Tech. Rep., 2014. [Online]. Available: http://www.shift2rail.org

[2] K. Guan et al., "On millimeter wave and THz mobile radio channel for smart rail mobility," *IEEE Trans. Veh. Technol.*, vol. 66, no. 7, pp. 5658–5674, 2016. doi: 10.1109/TVT.2016.2624504.

[3] M. Xiao et al., "Millimeter wave communications for future mobile networks," *IEEE J. Sel. Areas Commun.*, vol. 35, no. 9, pp. 1909–1935, Sept. 2017. doi: 10.1109/JSAC.2017.2719924.

[4] S. Mumtaz, J. M. Jornet, J. Aulin, W. H. Gerstacker, X. Dong, and B. Ai, "Terahertz communication for vehicular networks," *IEEE Trans. Veh. Technol.*, vol. 66, no. 7, pp. 5617–5625, July 2017. doi: 10.1109/TVT.2017.2712878.

[5] T. S. Rappaport, Y. Xing, G. R. MacCartney, A. F. Molisch, E. Mellios, and J. Zhang, "Overview of millimeter wave communications for fifth-generation (5G) wireless networks: With a focus on propagation models," *IEEE Trans. Antennas Propag.*, vol. 65, no. 12, pp. 6213–6230, Dec. 2017. doi: 10.1109/TAP.2017.2734243.

[6] J. Moreno, J. Riera, L. De Haro, and C. Rodriguez, "A survey on future railway radio communications services: Challenges and opportunities," *IEEE Commun. Mag.*, vol. 53, no. 10, pp. 62–68, Oct. 2015. doi: 10.1109/MCOM.2015.7295465.

[7] K. Guan, D. He, A. Hrovat, B. Ai, Z. Zhong, and T. Kürner, "Challenges and chances for smart rail mobility at mmWave and THz bands from the channels viewpoint," in *Proc. 15th Int. Conf. Intelligent Transport Systems (ITS) Telecommunications*, May 2017, pp. 1–5. doi: 10.1109/ITST.2017.7972207.

[8] P. Kyösti et al., "WINNER II channel models," D1.1.2 V1.1, 2007.

[9] D. He, B. Ai, K. Guan, L. Wang, Z. Zhong, and T. Kürner, "The design and applications of high-performance ray-tracing simulation platform for 5G and beyond wireless communications: A tutorial," *IEEE Commun. Surveys Tuts.*, to be published.

[10] K. Guan, Z. Zhong, B. Ai, and T. Kürner, "Deterministic propagation modeling for the realistic high-speed railway environment," in *Proc. 2013 IEEE 77th Vehicular Technology Conf. (VTC Spring)*, Dresden, Germany, pp. 1–5. doi: 10.1109/VTCSpring.2013.6692506.

[11] S. Priebe and T. Kürner, "Stochastic modeling of THz indoor radio channels," *IEEE Trans. Wireless Commun.*, vol. 12, no. 9, pp. 4445–4455, Sept. 2013. doi: 10.1109/TWC.2013.072313.121581.

[12] K. Guan et al., "Towards realistic high-speed train channels at 5G millimeter-wave band – part I: Paradigm, significance analysis, and scenario reconstruction," *IEEE Trans. Veh. Technol.*, vol. 67, no. 10, pp. 9129–9144, Aug. 2018. doi: 10.1109/TVT.2018.2865498.

[13] T. Abbas, J. Nuckelt, T. Kürner, T. Zemen, C. F. Mecklenbrauker, and F. Tufvesson, "Simulation and measurement-based vehicle-to-vehicle channel characterization: Accuracy and constraint analysis," *IEEE Trans. Antennas Propag.*, vol. 63, no. 7, pp. 3208–3218, 2015. doi: 10.1109/TAP.2015.2428280.

[14] K. Guan et al., "Towards realistic high-speed train channels at 5G millimeter-wave band – part II: Case study for paradigm implementation," *IEEE Trans. Veh. Technol.*, vol. 67, no. 10, pp. 9129–9144, Aug. 2018. doi: 10.1109/TVT.2018.2865530.

[15] K. Guan et al., "Scenario modules, ray-tracing simulations and analysis of millimetre wave and terahertz channels for smart rail mobility," *IET Microw., Antennas Propag.*, vol. 12, no. 4, pp. 501–508, Mar. 2018. doi: 10.1049/iet-map.2017.0591.

[16] X. Cheng et al., "Communicating in the real world: 3D MIMO," *IEEE Wireless Commun.*, vol. 21, no. 4, pp. 136–144, Aug. 2014. doi: 10.1109/MWC.2014.6882306.

[17] D. S. Baum, J. Hansen, and J. Salo, "An interim channel model for beyond-3G systems: Extending the 3GPP spatial channel model (SCM)," in *Proc. 2005 IEEE 61st Vehicular Technology Conf.*, vol. 5, pp. 3132–3136. doi: 10.1109/VETECS.2005.1543924.

[18] A. Ghazal et al., "A non-stationary IMT-advanced MIMO channel model for high-mobility wireless communication systems," *IEEE Trans. Wireless Commun.*, vol. 16, no. 4, pp. 2057–2068, Apr. 2017. doi: 10.1109/TWC.2016.2628795.

[19] X. Chen, "Throughput multiplexing efficiency for MIMO antenna characterization," *IEEE Antennas Wireless Propag. Lett.*, vol. 12, pp. 1208–1211, 2013. doi: 10.1109/LAWP.2013.2282952.

[20] J. Bian et al., "A WINNER+ based 3D non-stationary wideband MIMO channel model," *IEEE Trans. Wireless Commun.*, vol. 17, no. 3, pp. 1755–1767, Mar. 2018. doi: 10.1109/TWC.2017.2785249.

[21] J. Zhang, Y. Zhang, Y. Yu, R. Xu, Q. Zheng, and P. Zhang, "3D MIMO: How much does it meet our expectations observed from channel measurements?" *IEEE J. Sel. Areas Commun.*, vol. 35, no. 8, pp. 1887–1903, Aug. 2017. doi: 10.1109/JSAC.2017.2710758.

[22] S. Wu, C. Wang, e. M. Aggoune, M. M. Alwakeel, and X. You, "A general 3-D non-stationary 5G wireless channel model," *IEEE Trans. Commun.*, vol. 66, no. 7, pp. 3065–3078, July 2018. doi: 10.1109/TCOMM.2017.2779128.

[23] S. Jaeckel, L. Raschkowski, K. Börner, and L. Thiele, "QuaDRiGa: A 3-D multi-cell channel model with time evolution for enabling virtual field trials," *IEEE Trans. Antennas Propag.*, vol. 62, no. 6, pp. 3242–3256, June 2014. doi: 10.1109/TAP.2014.2310220.

[24] M. Zhu, G. Eriksson, and F. Tufvesson, "The COST 2100 channel model: Parameterization and validation based on outdoor MIMO measurements at 300 MHz," *IEEE Trans. Wireless Commun.*, vol. 12, no. 2, pp. 888–897, Feb. 2013. doi: 10.1109/TWC.2013.010413.120620.

[25] "Guidelines for evaluation of radio interface technologies for IMT-Advanced," Tech. Rep. ITU-R M.2135, 2008.

[26] X. Cheng, C.-X. Wang, B. Ai, and H. Aggoune, "Envelope level crossing rate and average fade duration of non-isotropic vehicle-to-vehicle Ricean fading channels," *IEEE Trans. Intell. Transp. Syst.*, vol. 15, no. 1, pp. 62–72, Feb. 2014. doi: 10.1109/TITS.2013.2274618.

[27] T. S. Rappaport, R. W. Heath Jr., R. C. Daniels, and J. N. Murdock, *Millimeter Wave Wireless Communications*. Englewood Cliffs, NJ: Prentice Hall, 2014.

[28] S. Hur et al., "Proposal on millimeter-wave channel modeling for 5G cellular system," *IEEE J. Sel. Topics Signal Process.*, vol. 10, no. 3, pp. 454–469, Apr. 2016. doi: 10.1109/JSTSP.2016.2527564.

[29] P. Series, "Propagation data and prediction methods for the planning of indoor radiocommunication systems and radio local area networks in the frequency range 900 MHz to 100 GHz," ITU-R, Geneva, Tech. Rep. ITU-R P.1238-7, Feb. 2012.

[30] J. Y. Yang et al., "A geometry-based stochastic channel model for the millimeter-wave band in a 3GPP high-speed train scenario," *IEEE Trans. Veh. Technol.*, vol. 67, no. 5, pp. 3853–3865, Jan. 2017. doi: 10.1109/TVT.2018.2795385.

[31] T. Kawanishi et al., "Proposal of a new working document of a draft new apt report on millimeter-wave band railway radiocommunication systems between train and trackside, and its work plan," National Inst. of Information and Communication Technol., Japan, Tech. Rep. AWG-20/INP-43, Sept. 2016.

[32] S. W. Choi et al., "Performance evaluation of millimeter-wave-based communication system in tunnels," in *Proc. 2015 IEEE Globecom Workshops (GC Wkshps)*, pp. 1–5. doi: 10.1109/GLOCOMW.2015.7414003.

[33] J. Kim, H. S. Chung, S. W. Choi, I. G. Kim, and Y. Han, "Mobile hotspot network enhancement system for high-speed railway communication," in *Proc. 2017 11th European Conf. Antennas and Propagation (EUCAP)*, pp. 2885–2889. doi: 10.23919/EuCAP.2017.7928212.

[34] Y. Liu, X. Fang, M. Xiao, and S. Mumtaz, "Decentralized beam pair selection in multi-beam millimeter-wave networks," *IEEE Trans. Commun.*, vol. 66, no. 6, pp. 2722–2737, June 2018. doi: 10.1109/TCOMM.2018.2800756.

[35] B. Peng and T. Kürner, "Three-dimensional angle of arrival estimation in dynamic indoor terahertz channels using a forward-backward algorithm," *IEEE Trans. Veh. Technol.*, vol. 66, no. 5, pp. 3798–3811, May 2017.